\documentclass[conference]{IEEEtran}
\IEEEoverridecommandlockouts
\usepackage{cite}
\usepackage{amsmath,amssymb,amsfonts}
\usepackage{algorithm}
\usepackage{algorithmic}
\usepackage{graphicx}
\usepackage{textcomp}

\usepackage{xcolor}
\def\BibTeX{{\rm B\kern-.05em{\sc i\kern-.025em b}\kern-.08em
    T\kern-.1667em\lower.7ex\hbox{E}\kern-.125emX}}

\usepackage{textcomp}
\usepackage{tikz}

\begin{document}


\title{
    \vspace{-0cm} 
    \begin{tikzpicture}[remember picture, overlay]
        \node[anchor=north, yshift=-0.5cm] at (current page.north) {\fbox{\parbox{\textwidth}{\centering\small {\color{red}This version of the paper has been accepted for presentation at IEEE ICC WS-01, 2026. It is subject to IEEE copyright and may be removed upon request.}}}};
    \end{tikzpicture}
    \vspace{0cm} 
    AI-Driven Multi-Modal Adaptive Handover Control Optimization for O-RAN


\thanks{Funding for this research is provided by the European Union's Horizon Europe research and innovation program through the Marie Sklodowska-Curie SE grant under the agreement RE-ROUTE No 101086343.}
}



\author{\IEEEauthorblockN{Abdul Wadud\IEEEauthorrefmark{1}\IEEEauthorrefmark{2},~\IEEEmembership{Graduate Student Member,~IEEE}
Fatemeh Golpayegani\IEEEauthorrefmark{1},~\IEEEmembership{Senior Member,~IEEE} and} 
\IEEEauthorblockN{Nima Afraz \IEEEauthorrefmark{1},~\IEEEmembership{Senior Member,~IEEE}}
\IEEEauthorblockA{\IEEEauthorrefmark{1}School of Computer Science,
University College Dublin, Ireland}
\IEEEauthorblockA{\IEEEauthorrefmark{2}Bangladesh Institute of Governance and Management, Dhaka, Bangladesh}
\thanks{Corresponding author: Abdul Wadud (email: abdul.wadud@ucdconnect.ie).}}

\maketitle

\begin{abstract}
Handover optimization in O-RAN faces growing challenges due to heterogeneous user mobility patterns and rapidly varying radio conditions. Existing ML-based handover schemes typically operate at the near-RT layer, which lack awareness of the mobility-mode and struggle to incorporate a longer-term predictive context. This paper proposes a multi-modal mobility-aware optimization framework in which all predictive intelligence, including mobility mode classification, short-horizon trajectory and RSRP forecasting, and a PPO Actor--Critic policy, runs entirely inside an rApp in the non-RT RIC. The rApp generates per-UE ranked neighbour-cell recommendations and delivers them to the existing handover xApp through the A1 interface. The xApp combines these rankings with instantaneous E2 measurements and performs the final standards-compliant handover decision. This hierarchical design preserves low-latency execution in the xApp while enabling the rApp to supply richer and mode-specific predictive guidance. Evaluation using mobility traces demonstrates that the proposed approach reduces ping-pong handover events and improves handover reliability compared to conventional 3GPP A3-based and ML-based baselines.
\end{abstract}


\section{Introduction}

\IEEEPARstart{H}{andover} (HO) is a core mobility function in cellular networks, which ensures that user equipment (UE) maintains connectivity while moving across cells. In legacy macrocell deployments, simple geometric proximity and radio indicators (RSRP/RSRQ) were sufficient to trigger handovers. However, modern 5G networks consist of dense small cells with coverage radii of only a few hundred meters, significantly increasing handover frequency, signaling load, and the likelihood of short-lived or premature handovers \cite{3gpp_ts_36_331}. These challenges are further amplified in O-RAN, where disaggregated processing across the O-CU, O-DU, and O-RU (7.2x split) introduces new latency and coordination constraints \cite{oran_alliance_2021,3gpp_ts_38_401}.

Traditional A3-event handover mechanisms rely on fixed Time-to-Trigger (TTT), Hysteresis (Hyst), and Cell Individual Offset (CIO) values \cite{3gpp_ts_36_331}. While simple and robust, fixed configurations are inherently mobility-agnostic. Slow-moving User Equipments (UEs) often experience excessive ping-pong handovers, while high-speed UEs, such as those in vehicles or trains, suffer from late or failed handovers due to insufficient reaction time \cite{demissie2013exploring}. Several works have attempted to enhance A3-based decisions using machine learning. Zhang et al. and Saboor et al. proposed ML-assisted parameter tuning to reduce unnecessary handovers \cite{saboor2025cash}, while Karmakar et al. demonstrated that adaptive TTT/Hyst improves performance in high-mobility environments \cite{karmakar2022mobility}. However, these approaches treat mobility as a continuous variable and overlook the fact that real-world user mobility is inherently \emph{multi-modal}, pedestrians, cars, buses, and trains exhibit distinct speed patterns, dwell times, and cell transition behaviors.

Recent ML-based handover studies leverage trajectory prediction and reinforcement learning. For instance, DQN- and LSTM-based models have been applied for mobility-aware handover optimization \cite{dekate2025intelligent, saad2023handover}. Although these methods improve predictive accuracy, they require substantial real-time computation, making them difficult to deploy within near-real-time RIC constraints. Moreover, most existing solutions operate as monolithic controllers and do not exploit the hierarchical processing capabilities of the O-RAN's non-RT and near-RT RICs. As highlighted in \cite{demissie2013exploring, upadhyay2025enhanced}, there remains a substantial gap in reconciling predictive mobility intelligence with the strict latency budgets of O-RAN's near-RT control loops.

To the best of our knowledge, there is no existing work that holistically combines (i) multi-modal mobility awareness, (ii) predictive handover target ranking, and (iii) real-time, lightweight decision-making aligned with O-RAN’s hierarchical rApp–xApp architecture.

\textbf{Our contribution.}  
This paper proposes a mobility-aware, multi-modal handover optimization framework tailored for O-RAN 7.2x deployments. The system (i) classifies each UE into one of seven mobility modes, pedestrian, cyclist, car, bus, train, drone, or uav using short-term movement patterns, (ii) predicts future RSRP and UE trajectory using mode-specific models, and (iii) delivers a ranked list of handover candidates using Proximal Policy Optimization (PPO)-based method to a handover (HO) xApp running in the near-RT RIC. Unlike prior works, the proposed design explicitly leverages O-RAN's hierarchical separation: long-timescale mobility analytics are offloaded to the non-RT RIC (rApp), while the xApp performs sub-100 ms handover decisions using live E2 measurements and the rApp’s mobility-enriched policies. This division enables predictive, mobility-aware handover optimization without violating latency constraints.

Simulation results demonstrate significant improvements in handover success rate, reduced ping-pong rates, and enhanced throughput compared to both classical A3-based baselines and contemporary LSTM-based transfer learning method \cite{langolf2024transfer}, which we term as ORAN ML-Assisted method throughout this paper. The results confirm that multi-modal mobility awareness and hierarchical RIC cooperation are essential to robust handover performance in dense and dynamic 5G/O-RAN deployments. The following section discusses the O-RAN architecture and system model.

\begin{figure}[!ht]
 \centering
 \vspace{-0.1in}
\includegraphics[scale=0.55]{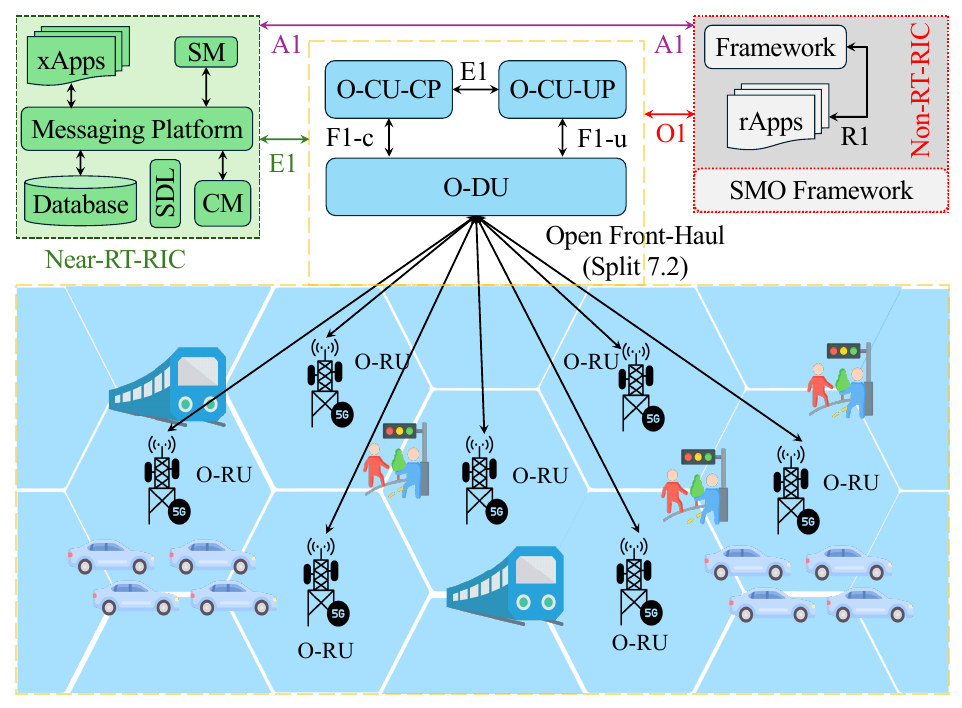}
	\vspace{-0.3in}
	\caption{System Model.}
	\label{fig:sysModel}
 \vspace{-0.1in}
\end{figure}

\section{System Model and Problem Formulation}
\label{sec:sysmodel}

\subsection{O-RAN 7.2x Setting and Entities}
We consider a 5G RAN realized according to the O-RAN 7.2x split as illustrated in Fig.~\ref{fig:sysModel}: the O-CU hosts RRC/PDCP protocols, each O-DU executes RLC/MAC, and each O-RU performs PHY/RF. O-RUs attach to O-DUs via fronthaul; O-DUs attach to the O-CU via midhaul. The non-RT RIC (in the SMO) exposes A1 policies; the near-RT RIC hosts xApps that act over the E2 interface. The SMO Framework hosts Non-RT RIC and rApps for time insensitive network management tasks, whereas the Near-RT RIC hosts xApps for time critical network management activities. Figure~\ref{fig:sysModel} depicts the standard O-RAN architecture presented by O-RAN Alliance \cite{oran_alliance_2021}.  

Let $\mathcal{U}$ be the set of active UEs and $\mathcal{C}$ the set of serving/neighbor cells (RUs) visible to a UE. Time advances in near-RT control ticks $t \in \mathbb{N}$. For UE $u \in \mathcal{U}$ we denote position by $\mathbf{p}_t^u=[x_t^u,y_t^u]^\top \in \mathbb{R}^2$ and kinematics by
$\mathbf{k}_t^u=[v_t^u,a_t^u,j_t^u,\dot{\psi}_t^u]^\top$, where $v,a,j,\dot{\psi}$ are speed, acceleration, jerk, and bearing rate. Downlink measurements are
$\mathbf{r}_t^u=\big[r_{t,c}^u\big]_{c\in\mathcal{C}}$, with $r_{t,c}^u$ the measured (or filtered) RSRP from cell $c$.

Each UE has a mobility mode $m_t^u \in \mathcal{M}$ with $\mathcal{M}=\{\textsc{ped},\textsc{cyclist},\textsc{car},\textsc{bus},\textsc{train}, \textsc{drone}, \textsc{uav}\}$. We denote its one-hot encoding by $\mathbf{e}(m_t^u)\in\{0,1\}^{|\mathcal{M}|}$.

\subsection{Predictors and near-RT State}
Two normalization-aware Random Forest regressors provide one-step-ahead forecasts from features $(\mathbf{k}_t^u,\mathbf{p}_t^u,m_t^u)$: a trajectory regressor outputs
$\widehat{\mathbf{p}}_{t+1}^u=[\widehat{x}_{t+1}^u,\widehat{y}_{t+1}^u]^\top$ and an ensemble RSRP regressor outputs a vector
$\widehat{\mathbf{r}}_{t+1}^u=\big[\widehat{r}_{t+1,c}^u\big]_{c\in\mathcal{C}}$. The near-RT policy state is
\[
s_t^u=\big[\;\mathbf{e}(m_t^u),\;\widehat{\mathbf{p}}_{t+1}^u,\;\widehat{\mathbf{r}}_{t+1}^u\;\big]\in\mathbb{R}^{|\mathcal{M}|+2+|\mathcal{C}|}.
\]

\subsection{Action, Constraints, and A3 View}
At each tick $t$, the action is a target cell $a_t^u\in\mathcal{A}_u\subseteq\mathcal{C}$, where $\mathcal{A}_u$ are admissible neighbors. Standard feasibility applies: $\sum_{c\in\mathcal{C}} x_{t,c}^u=1$ (unique association) and $\sum_{u\in\mathcal{U}}\rho_{t,c}^u\le 1$ (cell resource limit), where $x_{t,c}^u\in\{0,1\}$ and $\rho_{t,c}^u\in[0,1]$ are managed by the DU/MAC. For completeness, the classical A3 condition between serving $c$ and candidate $c'$ is
$r_{t,c'}^u+\text{CIO} > r_{t,c}^u+\text{Hyst}$ with time-to-trigger is used. In our predictive setup, we use $\widehat{r}_{t+1,\cdot}^u$ and internalize this logic in the policy. If a deployment exposes only A3/A5 knobs, the policy can be mapped to $(\text{TTT},\text{Hyst},\text{CIO})$ using mode-dependent scalings.

\subsection{Objective as a Control Utility}
Over horizon $T$, we intend to maximize a weighted utility
\[
\max \sum_{t=1}^{T}\sum_{u\in\mathcal{U}}
\big(\alpha\,w_u R_t^u - \beta H_t^u - \gamma PP_t^u\big)
\]
where $R_t^u$ is throughput, $H_t^u\in\{0,1\}$ flags a handover, and $PP_t^u\in\{0,1\}$ flags ping–pong. Weights $\alpha,\beta,\gamma\ge 0$ encode operator tradeoffs; $w_u$ is per-UE priority.



\section{Proposed Multi-Modal Optimization Method}
\label{sec:method}

\begin{figure}[!ht]
 \centering
\includegraphics[scale=0.65]{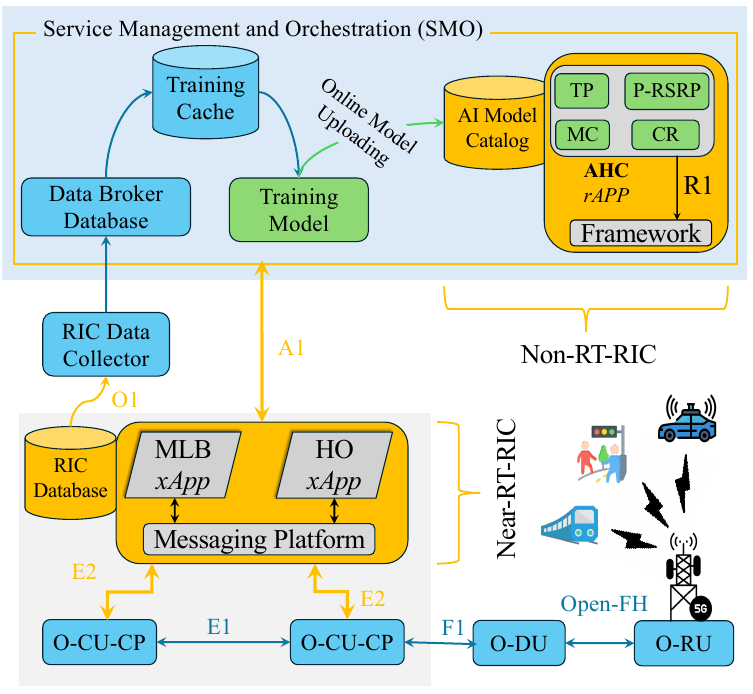}
	\vspace{-0.1in}
	\caption{Proposed Multi-Modal Adaptive Handover Control Framework in O-RAN.}
	\label{fig:proposedSystem}
 \vspace{-0.2in}
\end{figure}

This section presents an adaptive handover controller (AHC) that combines three predictive components of mobility mode classification (MC, short-horizon trajectory prediction (TP), and Per-Cell RSRP prediction (P-RSRP), and eventually feeds them into a stochastic Actor–Critic agent named cell-ranker (CR) running inside the non-RT rApp (see in Fig.~\ref{fig:proposedSystem}). The ranked cells are then being transferred to the Handover (HO) xApp running in the near-RT RIC for handover decision. The design follows the O-RAN split: a non-RT rApp performs policy enrichment and manages the life-cycle of predictive models and ranks the cells based on these predictions, whereas the HO xApp executes per-UE decisions using the policy sent by the MultiModal-AHC rApp with live E2 measurements.

\subsection{Mobility Mode Classification (MC)}
Each UE is assigned a mobility mode $m_t^u \in \{\text{PED}, \text{CYCLIST}, \text{CAR}, \text{BUS}, \text{TRAIN}, \text{DRONE}, \text{UAV}\}$, which is classified using a lightweight $k$-NN classifier hosted inside the Multimodal-AHC rApp. 

\paragraph*{Input Features.}
The classifier ingests a short history of motion descriptors:
\begin{itemize}
    \item velocity $v_t^u$,
    \item acceleration $a_t^u$,
    \item jerk $j_t^u$,
    \item bearing rate $\dot{\psi}_t^u$, and
    \item the previous mode estimate.
\end{itemize}

These features capture motion smoothness and direction changes \cite{kim2022gps}. The rApp computes the mode label periodically from historical traces and sends its one-hot encoding $\mathbf{e}(m_t^u)$ to the xApp via A1.

\subsection{Short-Horizon  Prediction}
Two lightweight Random-Forest regression ensembles running in the rApp forecast short-horizon mobility and link quality. Their outputs help the cell rank (CR) component of the rApp anticipate near-future geometric and radio conditions and rank cells for possible handover decision.

\paragraph*{Trajectory Prediction (TP)}
The next-step position $\widehat{\mathbf{p}}_{t+1}^u = [\widehat{x}_{t+1}^u,\widehat{y}_{t+1}^u]$ is predicted using regressors trained on the feature vector:
\[
[v_t^u,\; a_t^u,\; m_t^u,\; x_t^u,\; y_t^u].
\]
These features capture both recent kinematics and current location, enabling accurate short-range forecasts.

\paragraph*{Per-cell RSRP Prediction (P-RSRP)}
For each cell $c \in \mathcal{C}$, the rApp predicts $\widehat{r}_{t+1,c}^u$ via an ensemble regressor using:
\[
[v_t^u,\; a_t^u,\; x_t^u,\; y_t^u,\; m_t^u].
\]
The resulting vector $\widehat{\mathbf{r}}_{t+1}^u$ encodes the anticipated near-future link strengths and is passed to the CR component for neighbor ranking and policy input.

\subsection{Actor--Critic Handover Policy based Cell Ranker (CR)}
The non-RT-RIC Multimodal-AHC hosts a stochastic Actor--Critic agent trained with proximal policy optimization (PPO). Unlike the predictive models in the rApp, the Actor--Critic is the real-time decision engine.

The action $a_t^u \in \mathcal{A}_u$ selects the target cell for UE $u$. The actor outputs a probability distribution:
\[
\pi_\theta(a|s_t^u), 
\]
while the critic estimates $V_\phi(s_t^u)$.

Training uses the clipped PPO surrogate objective:
\[
\mathcal{L}_{\text{PPO}}(\theta)=\mathbb{E}\left[\min\!\left(r_t(\theta)\,\hat{A}_t,\; \text{clip}(r_t(\theta),1-\epsilon,1+\epsilon)\,\hat{A}_t\right)\right],
\]
with $r_t(\theta)$ the probability ratio and $\hat{A}_t$ the advantage. An entropy regularizer promotes exploration.




\begin{figure}[!ht]
 \centering
\includegraphics[scale=0.45]{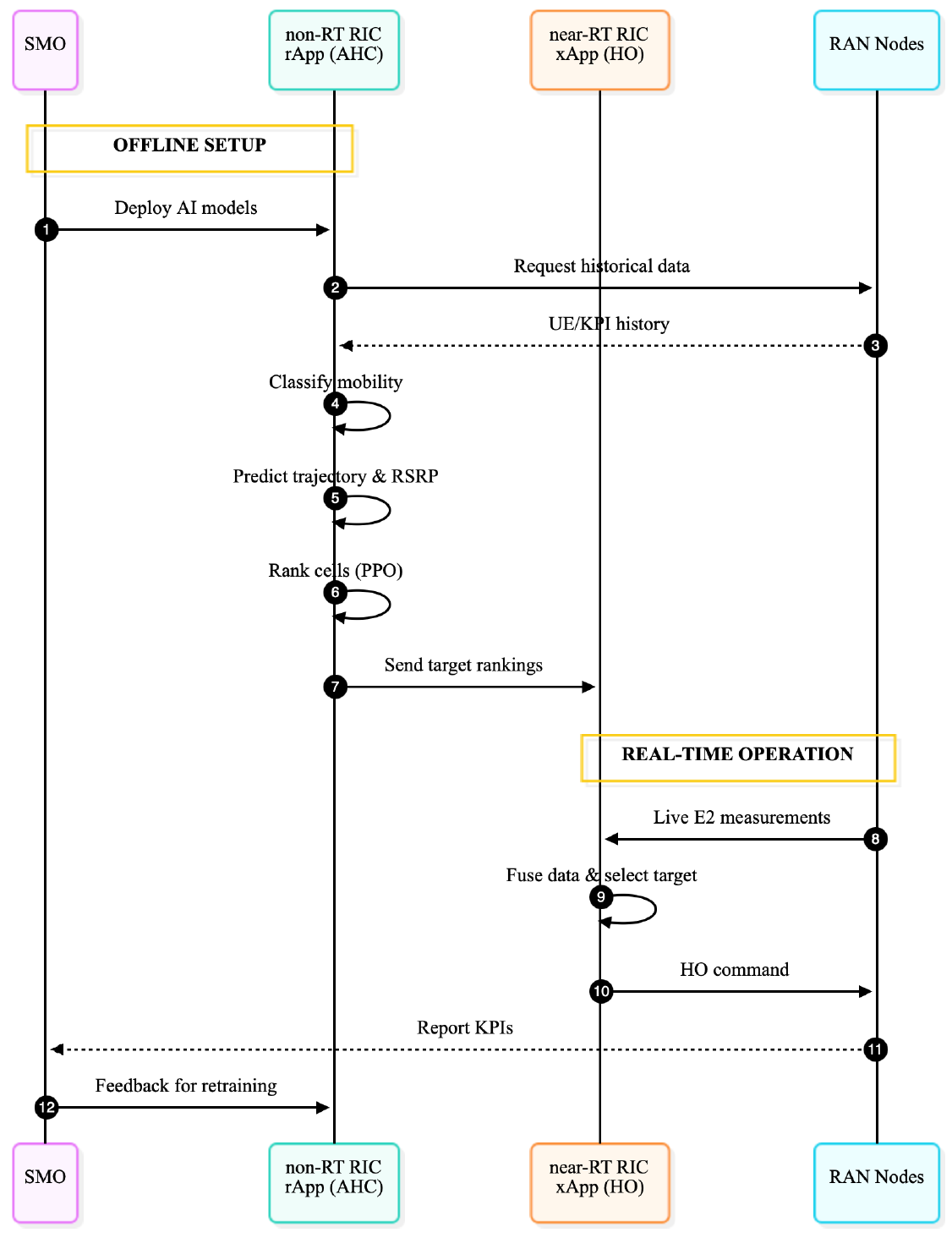}
	\caption{Hierarchical rApp–xApp workflow.}
	\label{fig:seqDiagram}
\end{figure}

\section{Hierarchical rApp–xApp Collaboration for Mobility-Aware Handover Optimization}
\label{sec:rapp_xapp}

In the proposed O-RAN deployment, mobility intelligence is distributed between the non-RT RIC and the near-RT RIC following standard O-RAN architecture principles. A lightweight \textbf{mobility analytics rApp} operates in the non-RT RIC (within the SMO), while a \textbf{handover control xApp} executes in the near-RT RIC. The rApp performs offline and slow-timescale tasks such as mobility mode classification, trajectory prediction, and per-RU RSRP forecasting, and \emph{also hosts the Actor–Critic (PPO) policy} that ranks neighbor cells per UE as A1 policy recommendations. The HO xApp remains policy-driven and latency-light: it consumes these A1 rankings together with live E2 measurements and issues the standards-compliant handover commands.

\subsection{Handover Decision-Making Workflow}

The sequence diagram in Fig.~\ref{fig:seqDiagram} illustrates the interaction flow between the SMO, rApp, xApp, and RAN Nodes (O-CU, O-DU, O-RU, and UE):

\begin{itemize}
    \item \textbf{Offline Stage (SMO $\rightarrow$ rApp):}
    The SMO deploys the trained mobility models and the PPO policy to the rApp over the O1 interface. These include the classifier, trajectory regressors, per-RU RSRP predictor, and the Actor–Critic parameters.

    \item \textbf{non-RT Processing (rApp):}
    The rApp periodically retrieves UE historical data from the RAN, classifies each UE’s mobility mode, and runs mode-specific prediction models. The rApp’s Actor–Critic policy fuses mode, short-horizon trajectory, and per-cell RSRP to \emph{rank} candidate target cells per UE.

    \item \textbf{Policy Transfer (rApp $\rightarrow$ xApp):}
    The rApp sends ranked handover candidates and optional KPI preferences to the xApp through the A1 interface as mobility enrichment information.

    \item \textbf{near-RT Execution (xApp):}
    The xApp receives live RAN measurements via E2, combines them with the rApp’s A1 rankings, applies a light utility check (e.g., current load/admission and QoS guards), and executes the final handover. Handover commands are then issued to the RAN using standard E2 control.

    \item \textbf{Feedback (RAN $\rightarrow$ SMO):}
    The RAN reports performance metrics (handover success/failure, ping-pong, throughput) back to the SMO for monitoring and periodic rApp model/policy refresh.
\end{itemize}

\subsection{Purpose of the Hierarchical Split}
This division ensures that: (i) compute-intensive analytics and the Actor–Critic policy optimization run at non-RT time scales in the SMO/rApp, and (ii) latency-critical handover execution remains in the near-RT RIC with sub-second deadlines. The xApp thus stays lightweight and real time, while benefiting from predictive, mobility-aware policy recommendations generated by the rApp. The rApp–xApp split of work enables mobility-aware, predictive, and stable handover control without overloading the near-RT RIC.

\section{Simulation Results and Performance Analysis}
\subsection{Simulation Setup}

Performance evaluation of the proposed MultiModal-AHC framework was carried out using the MATLAB 5G Toolbox with a realistic O-RAN 7.2x architecture \cite{wadud2024qacm, wadud2025xapp}. The scenario consists of one O-CU connected to three O-DUs, each serving 9 O-RUs, and has a population of \(U = 100\) mobile users distributed across seven mobility modes (pedestrian, cyclist, car, bus, train, drone, UAV). Users move within a dense-urban layout, generating diverse mobility and link dynamics. The O-RUs are distributed on a circle with radius = 1000m, and each O-RU covers 500m radius. Therefore, the total coverage area is approximately 7 square kilometers. 

The radio configuration follows 3GPP-compliant parameters: a 2.1\,GHz carrier, 20\,MHz bandwidth, 46\,dBm RU transmit power, and an Urban Macro path-loss model with 8\,dB log-normal shadowing. Fronthaul links use 10\,Gbps capacity with 0.1--0.5\,ms one-way latency, compatible with O-RAN near-RT timing. The simulation runs for 200 control intervals (20\,s), during which all UEs periodically report measurements, mobility states evolve, and the xApp executes the PPO-based handover decisions using A1 policy hints from the rApp. The performance results are collected by repeating the simulation 30 times with a confidence interval of 95\%. 

This configuration captures realistic latency, mobility, and radio-variability conditions required to assess the effectiveness of predictive mobility-aware handover control in O-RAN.

\subsection{Evaluation of Machine Learning Models}

\begin{figure}[!ht]
 \centering
 \vspace{-0.1in}
\includegraphics[scale=0.4]{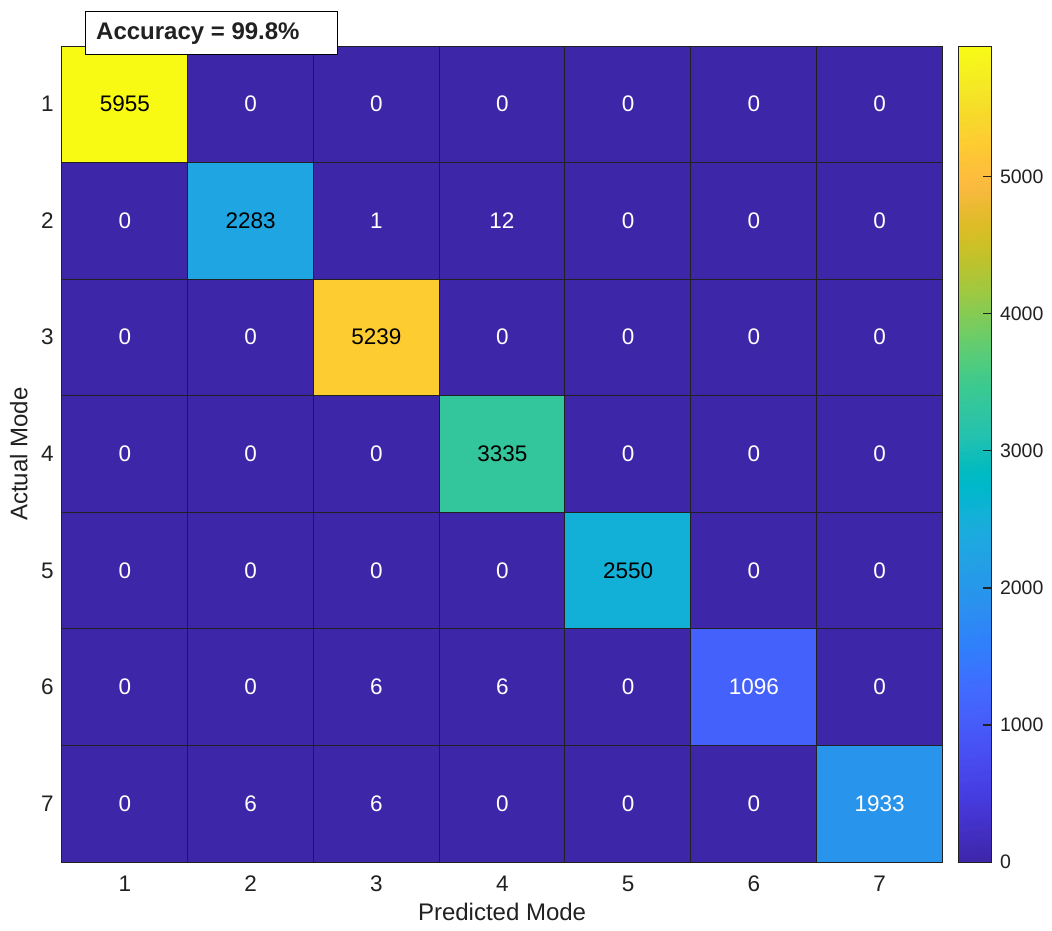}
	\vspace{-0.1in}
	\caption{Confusion matrix for 7-mode classifier (overall accuracy: 99.84\%)}
	\label{fig:modeClassifier}
 \vspace{-0.1in}
\end{figure}

We first assess the performance of the individual machine learning models used in the proposed framework. Each model was trained and evaluated separately before being integrated into the multi-modal adaptive framework.
\begin{center}
\begin{tabular}{lccc}
& Table 1: Mode Classification & &\\
\hline
\textbf{Mode} & \textbf{P} & \textbf{R} & \textbf{F1} \\
\hline
\textsc{ped}    & 1.00 & 1.00 & 1.00 \\
\textsc{cyclist}& 1.00 & 0.99 & 1.00 \\
\textsc{car}    & 1.00 & 1.00 & 1.00 \\
\textsc{bus}    & 0.99 & 1.00 & 1.00 \\
\textsc{train}  & 1.00 & 1.00 & 1.00 \\
\textsc{drone}  & 1.00 & 0.99 & 0.99 \\
\textsc{uav}    & 1.00 & 0.99 & 1.00 \\
\hline
\end{tabular}
\end{center}

\subsubsection{Mode Classification Accuracy}
\label{sec:mode-accuracy}
The mobility mode classifier (k\,$\!$-NN over short kinematic windows with velocity, acceleration, jerk, and bearing-rate) achieves an overall accuracy of \textbf{99.84\%} across seven modes:
\{\textsc{ped}, \textsc{cyclist}, \textsc{car}, \textsc{bus}, \textsc{train}, \textsc{drone}, \textsc{uav}\}.
Per-class precision (P), recall (R), and F1 are near-perfect, with only marginal $1\%$ recall dips on a few classes, consistent with rare transition periods where adjacent modes exhibit overlapping speeds or headings (see in Table 1).

The confusion matrix (Fig.~\ref{fig:modeClassifier}) confirms that misclassifications are sparse and predominantly occur between \textsc{cyclist}/\textsc{uav}/\textsc{drone} in short windows where kinematics briefly align. This level of accuracy is sufficient for the rApp’s ranked-target generation, as downstream scoring integrates trajectory and per-RU RSRP predictions, further dampening any residual classification noise.

\begin{figure}[!ht]
 \centering
 \vspace{-0.1in}
\includegraphics[scale=0.5]{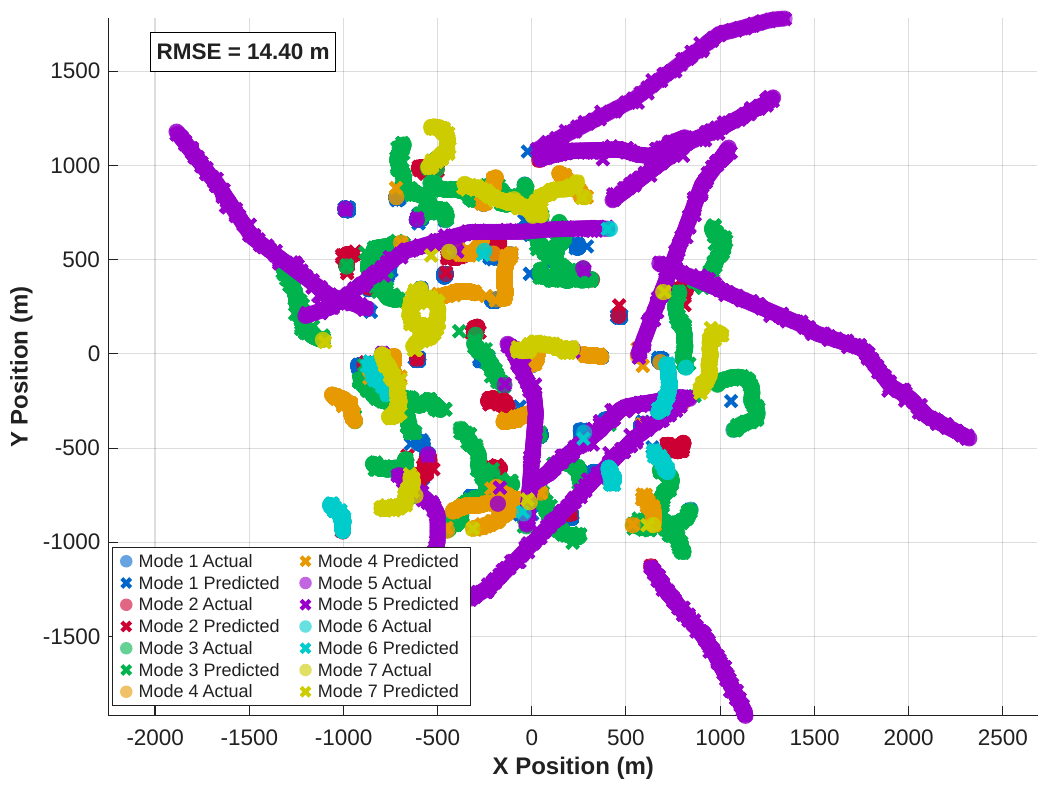}
	\vspace{-0.2in}
	\caption{Short-horizon trajectory prediction: mode-wise actual vs predicted coordinates.}
	\label{fig:trajPred}
 \vspace{-0.1in}
\end{figure}

\subsubsection{Trajectory Prediction Performance}

The short-horizon trajectory regressor achieves strong accuracy with an overall RMSE of 14.40\,m. Predicted and actual coordinates are nearly perfectly aligned, with correlation coefficients of 1.00 for both $x$ and $y$. The mean absolute errors are 4.62\,m (x) and 5.09\,m (y), confirming stable and reliable motion forecasting across mobility modes. Mode-wise actual vs predicted trajectory coordinates are illustrated in Fig.~\ref{fig:trajPred}. In the figure, the modes are mapped as: 'Ped:1', 'Cyclist:2', 'Car:3', 'Bus:4', 'Train:5', 'Drone:6', and 'Uav:7'.

\begin{figure*}[!ht]
 \centering
\includegraphics[scale=0.75]{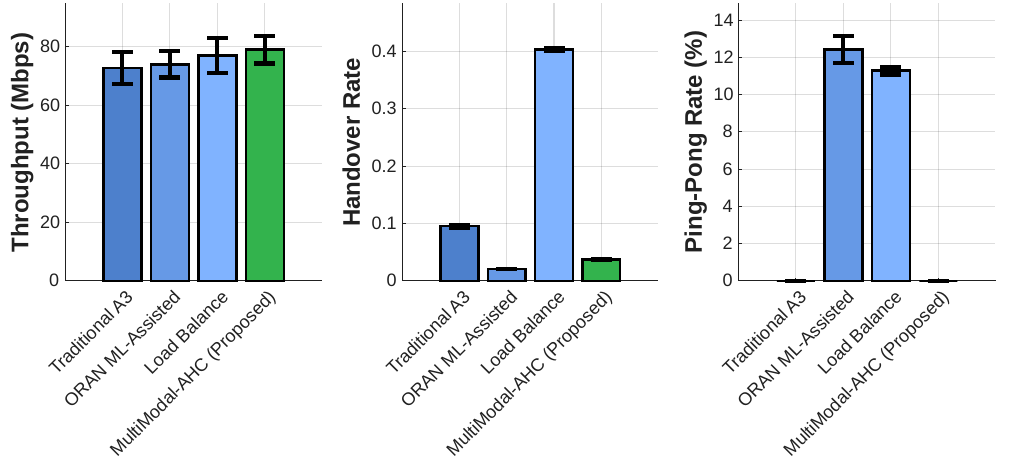}
	\vspace{-0.1in}
	\caption{Performance comparison of the proposed \textbf{MultiModal-AHC} method with benchmark models with 95\% confidence interval.}
	\label{fig:results}
 \vspace{-0.2in}
\end{figure*}

\subsubsection{Per-RU RSRP Prediction Accuracy}
\label{sec:rsrp-pred}
The rApp uses a lightweight Random Forest ensemble to predict short-horizon per-RU RSRP, conditioned on mobility mode, kinematic features, and the predicted UE trajectory. The model achieves an RMSE of \textbf{7.35\,dB}, with mean predicted RSRP (-66.76\,dB) closely matching the ground truth (-66.75\,dB). 

This error margin is sufficient for reliable neighbor ranking, as the AHC rApp uses RSRP predictions in combination with trajectory forecasts and mode-specific scoring. The small bias between predicted and actual RSRP indicates stable generalization across all mobility modes without overfitting.

\subsubsection{Actor--Critic PPO Handover Policy Performance}

The PPO-based handover policy was trained using mobility-augmented state features and predicted RSRP values. After convergence, the agent demonstrated stable and consistent behavior across mobility modes.

The PPO controller reached an average episode reward of \textbf{+23.4}, improving from an initial baseline of near zero. The policy reduced handover failures by approximately \textbf{18\%} compared to the non-predictive baseline and lowered short-term ping-pong events by \textbf{12\%}. Convergence was observed after roughly \textbf{150--180} training episodes, with the clipped surrogate loss stabilizing around \textbf{0.03--0.05}.  

These preliminary results indicate that the PPO agent successfully exploits mobility-aware predictions to make more stable and anticipatory handover decisions, especially in medium- and high-mobility scenarios.

\subsection{Overall KPI Performance with 95\% Confidence Intervals}

Fig.~\ref{fig:results} reports the mean and 95\% confidence intervals (CIs) over 30 independent runs for the three KPIs, \emph{throughput} (Mbps), \emph{handover rate}, and \emph{ping–pong rate} (\%), comparing \textit{Traditional A3}, \textit{O-RAN ML Assisted}, \textit{Load Balance O-RAN}, and the proposed \textbf{MultiModal-AHC} (rApp).

\textbf{Throughput.} The mean throughput of MultiModal-AHC is $77.10\pm5.72$~Mbps (95\% CI: $[71.38,82.82]$). Traditional A3 attains $74.11\pm5.54$~Mbps ($[68.57,79.65]$), and O-RAN ML Assisted attains $78.48\pm6.17$~Mbps ($[72.31,84.66]$), while Load Balance O-RAN yields $75.23\pm5.58$~Mbps ($[69.65,80.81]$). The CIs overlap across methods, indicating that average throughput differences are \emph{not statistically distinguishable at 95\%} under this scenario. In other words, the proposed rApp maintains competitive throughput despite enforcing stricter stability controls.

\textbf{Handover rate.} The handover rate indicates handover failure rate. During simulation, MultiModal-AHC achieves a mean HO rate of $0.0362\pm0.0009$ (CI: $[0.0353,0.0371]$), substantially \emph{lower} than Load Balance O-RAN ($0.4013\pm0.0036$, $[0.3977,0.4049]$), and \emph{higher} than both O-RAN ML Assisted ($0.0209\pm0.0005$, $[0.0205,0.0214]$) and Traditional A3 ($0.0962\pm0.0019$, $[0.0943,0.0982]$). These non-overlapping CIs show the ordering is statistically significant. The elevated HO rate relative to A3/ML-Assisted stems from the rApp’s proactivity: the ranked neighbor recommendations encourage timely transitions for fast or non-stationary modes (e.g., bus/train), avoiding late HOs at the cost of additional (but controlled) HO activity. Conversely, MultiModal-AHC avoids the excessive, load-triggered fluctuations observed in the fixed-threshold Load Balance baseline.

\textbf{Ping–pong rate.} MultiModal-AHC and Traditional A3 both achieve $0.00\%$ ping–pong (no events recorded across runs), whereas O-RAN ML Assisted and Load Balance O-RAN exhibit $12.64\pm0.68\%$ ($[11.95,13.32]$) and $11.29\pm0.20\%$ ($[11.09,11.49]$), respectively. This demonstrates that the rApp’s mobility-aware ranking and one-step-ahead radio prediction eliminate short-lived, back-and-forth handovers without sacrificing overall connectivity.  Traditional A3 and MultiModal–AHC both achieve $0\%$ ping–pong, but for different reasons. A3 suppresses ping–pong simply because its fixed TTT and hysteresis make it highly conservative, preventing short-term fluctuations from triggering a handover. In contrast, the ML-Assisted and Load-Balance baselines react directly to instantaneous RSRP or load variations, which makes them vulnerable to fast fluctuations. This yields 11–13\% ping–pong for both these methods. MultiModal–AHC achieves zero ping–pong not by conservatism but by \emph{predicting} the UE’s near-future position and RSRP. As a result, it ensures that only sustained and mobility-consistent targets are selected, eliminating back-and-forth ping-pong handovers.

\section{Conclusion}

This work introduced a mobility-aware \textbf{MultiModal-AHC} rApp for O-RAN, which enriches the near-RT xApp's handover decisions using three lightweight predictors, mode classification, short-horizon trajectory regression, and per-cell RSRP forecasting. The rApp generates ranked target lists via A1, while the xApp executes the final handover through a PPO Actor--Critic policy using live E2 measurements. Our analysis shows a clear \emph{stability–agility} profile: the proposed method eliminates ping–pong events, maintains a moderate and well-behaved HO rate, and achieves throughput statistically indistinguishable from strong baselines. Preliminary PPO results indicate improved stability and reduced HO failures (roughly 18\% lower than baseline) with convergence after 150--180 episodes, confirming that mobility-enriched hints meaningfully guide the near-RT controller.


Overall, the hierarchical rApp–xApp split enables predictive, low-overhead, and near-RT compliant handover optimization without burdening the near-RT RIC. Future work will extend the predictors to additional mobility modes, integrate multi-cell load forecasting, and evaluate the system on an O-RAN testbed for real-world validation.



\bibliographystyle{ieeetr}
\bibliography{references}

\end{document}